\documentclass[prb,twocolumn,showpacs,amsmath,amssymb,preprintnumbers,superscriptaddress]{revtex4}
\usepackage{dcolumn}
\usepackage{bm}
\usepackage{graphicx}
\usepackage{color}

\begin{document}
\title{Magnetic and transport properties in pyrochlore iridates (Y$_{1-x}$Pr$_x$)$_2$Ir$_2$O$_7$: The role of $f$-$d$ exchange interaction and $d$-$p$ orbital hybridization.}
\author{Harish Kumar}\affiliation{School of Physical Sciences, Jawaharlal Nehru University, New Delhi - 110067, India.}
\author{K. C. Kharkwal}\affiliation{School of Physical Sciences, Jawaharlal Nehru University, New Delhi - 110067, India.}
\author{Kranti Kumar}\affiliation{UGC-DAE Consortium for Scientific Research, Indore - 452001, India.}
\author{K. Asokan}\affiliation{Materials Science Division, Inter University Accelerator Centre, New Delhi- 110 067, India.}
\author{A. Banerjee}\affiliation{UGC-DAE Consortium for Scientific Research, Indore - 452001, India.}
\author{A. K. Pramanik}\email{akpramanik@mail.jnu.ac.in}\affiliation{School of Physical Sciences, Jawaharlal Nehru University, New Delhi - 110067, India.}

\begin{abstract}
The $f$-$d$ magnetic exchange interaction is considered to be a key ingredient for many exotic topological phases in pyrochlore iridates. Here, we have investigated the evolution of structural, magnetic and electronic properties in doped pyrochlore iridate, (Y$_{1-x}$Pr$_x$)$_2$Ir$_2$O$_7$. Apart from geometrical frustration, pyrochlore iridates are well known for its active spin-orbit coupling effect. The substitution of Pr$^{3+}$ (4$f^2$) for the nonmagnetic Y$^{3+}$ (4$d^0$) acts as a magnetic doping, which provides an ideal platform to study $f$-$d$ exchange interaction without altering the Ir-sublattice. With Pr substitution, system retains its original cubic structural symmetry but the local structural parameters show an evolution with the doping concentration $x$. The robust magnetic-insulating state in Y$_2$Ir$_2$O$_7$ is drastically weakened, while Pr$_2$Ir$_2$O$_7$ ($x$ = 1.0) shows a paramagnetic-metallic behavior. A metal-insulator transition is observed for $x$ = 0.8 sample. This evolution of magnetic and electronic properties are believed to be induced by an exchange interaction between localized Pr-4$f$ and itinerant Ir-5$d$ electrons as well as by an increased hybridization between Ir-$t_{2g}$ and (basal) O-$p$ orbitals as observed in XAS study. The resistivity in insulating materials follows a power-law behavior with a decreasing exponent with $x$. A negative magnetoresistance is observed for present series of samples at low temperature and where the magnetoresistance shows a quadratic field dependence at higher fields.
\end{abstract}

\pacs{75.47.Lx, 75.40.Cx, 74.25.Fy, 78.70.Dm}

\maketitle
\section {Introduction}
Geometrical frustration is inherent to pyrochlore systems which attaches many interesting properties.\cite{gingras,yoshii,gardner1,nakatsuji,bramwell,fukazawa1} The frustration mainly arises due to interpenetrating tetrahedra where the magnetic ions sitting at the vertices of tetrahedra introduce magnetic frustration for an antiferromagnetic (AFM) interactions. Ir based pyrochlores (A$_2$Ir$_2$O$_7$, A is trivalent rare earth elements) have, recently, drawn significant interest due to its novel topological phases of matter.\cite{william,gard,pesin} Apart from geometrical frustration, pyrochlore iridates share a comparable energy scale between spin-orbit coupling (SOC), electronic correlation (\textit{U}) and crystal field effect (CFE) due to 5\textit{d} based heavy Ir atoms (Z = 77). This complex interplay between these energies is believed to be a key ingredient for many interesting properties in this class of materials.

The size of A-site cations has an important role on physical properties of pyrochlore iridates giving rise from magnetic insulating to nonmagnetic complex metallic phases with increasing ionic radii.\cite{Matsuhira} The magnetic nature of A-ions further introduces complication in magnetic behavior due to possibility of \textit{f-d} exchange interactions between Ir- and A-ions.\cite{chen} The Y$_2$Ir$_2$O$_7$ is an important member of pyrochlore iridate family which has nonmagnetic Y$^{3+}$ at A-site, and shows an insulating behavior and AFM type magnetic transition with ordering temperature T$_{N}$ $\sim$ 160 K.\cite{shapiro,disseler,taira,fukazawa,zhu,soda,harish,kumar,hkumar,hari,vinod} Given that magnetically inactive Y$^{3+}$, this magnetic ordering is only associated with Ir sublattice. The low temperature magnetic state in Y$_2$Ir$_2$O$_7$ has been investigated using different microscopic tools. For instance, muon spin relaxation ($\mu$SR) measurements have shown spontaneous oscillation in muon symmetry, indicating a magnetically long-range ordering at low temperature.\cite{disseler,disseler1} The neutron powder diffraction (NPD) measurements, on the other hand, though have not ruled out the possibility of long-range magnetic ordering but because of iridium is strong absorber of neutrons, the NPD measurements remain largely insensitive for Ir containing materials.\cite{shapiro} Recently, using magnetic relaxation measurements we have shown a nonequilibrium magnetic ground state in Y$_2$Ir$_2$O$_7$ and its doped samples.\cite{harish,kumar,hkumar} Theoretically, this material is described as a possible candidate of Weyl-type semimetal with a noncoplanar AFM structure.\cite{wan} The Pr$_2$Ir$_2$O$_7$, on other hand, is another interesting compound in pyrochlore iridates which has magnetic and comparatively large A-site ion, Pr$^{3+}$ (4$f^{2}$). In other sense, this material has two active (Pr and Ir) sublattices. This material is metal and shows a nonmagnetic character, even though an AFM-type RKKY interaction of energy scale $\left|T^{*}\right|$ = 20 K between Pr-4\textit{f} moments has been observed which is mediated through Ir-5\textit{d} delocalized electrons.\cite{nakatsuji,Tokiwa} The AFM interaction is suppressed due to screening of 4\textit{f} moments through Kondo effect which decreases the Weiss temperature down to $\theta_P$ = 1.7 K. The Pr$_2$Ir$_2$O$_7$ is shown to exhibit spin liquid behavior where the system does not show any trace of magnetic ordering, even down to 70 mK.\cite{nakatsuji} Instead, a partial spin freezing of Pr-4\textit{f} moments is observed at 120 mK.\cite{nakatsuji} A large value of calculated frustration parameter (\textit{f} = 170) indicates high frustration in this material. Further, an unconventional anomalous Hall effect has been observed below 1 K in Pr$_2$Ir$_2$O$_7$ which is explained with the spin-chirality effect in Ir-5\textit{d} electrons due to noncoplaner spin structure of Pr spins.\cite{machida}

In present study, we have attempted to understand the evolution of magnetic and transport properties in pyrochlore iridate (Y$_{1-x}$Pr$_x$)$_2$Ir$_2$O$_7$ where Y$^{3+}$ is progressively replaced with Pr$^{3+}$. While this substitution induces \textit{f-d} interaction between Pr-4\textit{f} and Ir-5\textit{d} localized spins but other basic interactions such as, SOC and \textit{U} remain largely unaffected. The \textit{f-d} exchange interaction along with correlation effect has been theoretically shown to induce exotic topological electronic phases in pyrochlore iridates.\cite{chen} However, Pr$^{3+}$ being a large ion, is likely to induce local structural modification which as a result will influence the Ir-O interactions and hybridization. It is very important to investigate the evolution of magnetic and transport properties in present series, as Y$_2$Ir$_2$O$_7$ shows a high ordering temperature ($T_{N}$ $\sim$ 160 K) with a nonmagnetic A-site atom whereas Pr$_2$Ir$_2$O$_7$ avoids long-range magnetic ordering in spite of having magnetic ion sitting at A-site. This draws further attention as recently, a new type of quantum criticality, based on an interplay between SOC and $U$, has been proposed in pyrochlore iridates with a model system of (Y$_{1-x}$Pr$_x$)$_2$Ir$_2$O$_7$.\cite{lucile}

Our results show that while original cubic structural symmetry is retained in (Y$_{1-x}$Pr$_x$)$_2$Ir$_2$O$_7$ series but a minor modification in structural parameters is observed. The magnetic ordering temperature progressively decreases with Pr substitution where Pr$_2$Ir$_2$O$_7$ shows complete paramagnetic (PM) behavior. Similarly, this system shows a progressive metallic behavior exhibiting a metal-to-insulator transition for $x$ = 0.8 samples, where an enhanced Ir-O hybridization with an increasing Pr concentration plays a crucial role.

\section {Experimental Details}
Series of polycrystalline samples of (Y$_{1-x}$Pr$_x$)$_2$Ir$_2$O$_7$ with $x$ = 0.0, 0.05, 0.1, 0.2, 0.4, 0.6, 0.8 and 1.0 have been prepared using standard solid state route.\cite{harish,kumar,hkumar,hari} Ingredient powder materials Y$_2$O$_3$, Pr$_6$O$_{11}$, and IrO$_2$ with a phase purity more than 99.99\% (M/s Sigma-Aldrich) are mixed in stoichiometric ratio and ground well. The powder materials Y$_2$O$_3$, Pr$_6$O$_{11}$, have been given pre-heat treatment at 800$^o$C for around 8 hours to remove any residual atmospheric moisture. The well ground mixture is subsequently pelletized and sintered in air at temperature range between 1000 - 1160$^o$C about 19 days with several intermediate grindings. The crystal structure and phase purity of these materials have been checked with powder x-ray diffraction (XRD) using a Rigaku made diffractometer attached with CuK$_\alpha$ radiation. XRD data are collected in the range of 2$\theta$ = 10 - 90$^o$ at a step of $\Delta 2\theta$ = 0.02$^o$. The XRD data have been analyzed using Reitveld refinement program (FULLPROF) by Rodriguez Carvajal.\cite{carva} X-ray absorption measurements were performed at beamlines of 20A1 HSGM and 17C1 Wiggler of NSRRC, Taiwan as per standard procedure and data analysis. DC Magnetization (\textit{M}) measurements have been carried out with a vibrating sample magnetometer (PPMS, Quantum Design). Electrical transport properties have been measured using an integrated system from NanoMagnetics Instruments and Cryomagnetics, Inc.

\begin{figure}
	\centering
		\includegraphics[width=6.5cm]{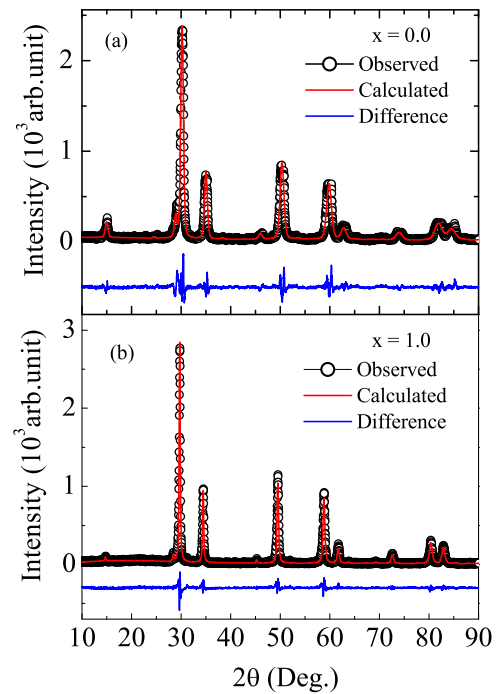}
	\caption{Room temperature XRD pattern along with Rietveld analysis have been shown for (Y$_{1-x}$Pr$_x$)$_2$Ir$_2$O$_7$ series with (a) $x$ = 0.0 and (b) $x$ = 1.0.}
	\label{fig:Fig1}
\end{figure}

\begin{figure}
	\centering
		\includegraphics[width=7.5cm]{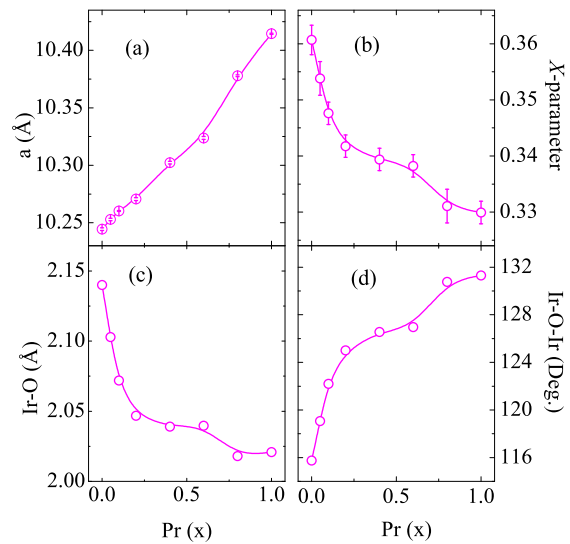}
	\caption{(a) Lattice constant $a$, (b) $X$- parameter of O atom, (c) Ir-O bond length and (d) Ir-O-Ir bond angle are shown as a function of Pr substitution for (Y$_{1-x}$Pr$_x$)$_2$Ir$_2$O$_7$ series. Lines are guide to eyes.}
	\label{fig:Fig2}
\end{figure}

\section{Results and Discussions}
\subsection {Structural analysis}
The XRD pattern along with Rietveld refinement are shown for two end members of the series i.e., Y$_2$Ir$_2$O$_7$ and Pr$_2$Ir$_2$O$_7$ in Figs. 1a and 1b, respectively. The XRD data for $x$ = 0.0 in Fig. 1a are in fact shown in our previous reports,\cite{harish,kumar} and here has been shown again for comparison in present series. With Pr substitution, no significant modification is observed in the XRD pattern in terms of peak position or arising of new peak(s). The Rietveld refinement of XRD data for Y$_2$Ir$_2$O$_7$ and whole series shows the materials crystallize in cubic \textit{Fd$\bar{3}$m} symmetry. We obtain a reasonable good fitting of Rietveld refinement, obtaining goodness of fit (GOF = $R_{wp}/R_{exp}$) values of the fitting between 1.5 - 2.0. We find no structural phase transition with Pr substitution, however, the structural parameters show an evolution with doping $x$ as shown in Fig. 2. Considering a slight mismatch in ionic radii between Y$^{3+}$ (1.019 \AA) and Pr$^{3+}$ (1.126 \AA), changes in structural parameters are expected. The unit cell parameter ($a$) for the parent compound Y$_2$Ir$_2$O$_7$ is found to be $a$ = 10.244(1) \AA \space which increases almost linearly with a slope about 0.17 upon substitution of Pr. For Pr$_2$Ir$_2$O$_7$ ($x$ = 1), $a$ is found to be 10.4145 \AA \space (Fig.2a) and change in $a$ mostly follows Vegerd’s law. The XRD pattern as well as the lattice parameter for both Y$_2$Ir$_2$O$_7$ and Pr$_2$Ir$_2$O$_7$ match well with the previous reports,\cite{shapiro,disseler,zhu,harish,kumar,kimura,millican}. The change in $a$ is about 1.6\% over whole series which can be explained due to larger ionic size of Pr$^{3+}$ compared to Y$^{3+}$ ions. Here, we add that the structural parameters in present series are obtained from Rietveld analysis of XRD data which is basically a bulk measurement and remains largely insensitive to local variation. We find that all the samples are ‘uniform’ in terms of chemical phase purity, where the Rietveld analysis imply an absence of any unwanted chemically impure phase. Given that Y$^{3+}$ and Pr$^{3+}$ have difference in ionic radii and they randomly populate the $A$-site, there would be a local disorder which cannot be understood from XRD data.

\begin{figure}
	\centering
		\includegraphics[width=7.5cm]{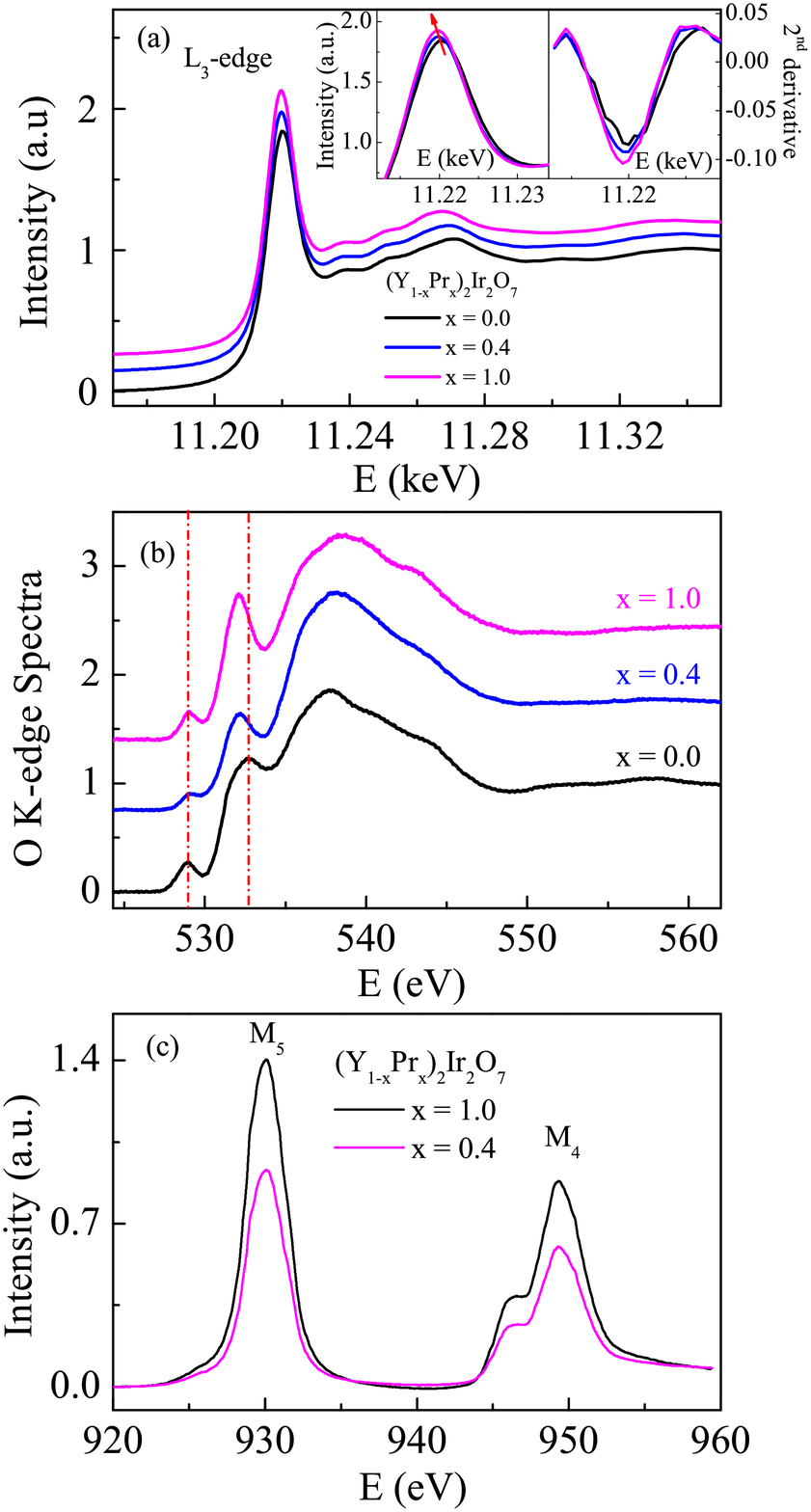}
	\caption{(a) XAS spectra at $L_3$ edge have been shown for (Y$_{1-x}$Pr$_x$)$_2$Ir$_2$O$_7$ series with $x$ = 0.0, 0.4 and 1.0. XAS data for $x$ = 0.4 and 1.0 are shifted vertically for clarity. Left and right inset show a magnified picture of $L_3$ absorption edge across 11220 eV and a second derivative of peaks, respectively. (b) Low energy O K-edge XAS spectra are shown for same $x$ = 0.0, 0.4 and 1.0 materials. (c) Normalized $M_{4}$ and $M_5$-edge XAS spectra are shown as a function of energy for $x$ = 1.0 and 0.4.}
	\label{fig:Fig3}
\end{figure}

In pyrochlore structure, there are mainly two variable parameters such as, lattice constant (a) and the  parameter associated with the oxygen position. This $X$-parameter plays a significant role in determining the electronic and magnetic properties in this class of materials.\cite{william} For instance, it decides the structural organization of IrO$_{6}$ octahedra which is a crucial in pyrochlore systems to decide its structural stability and physical properties. Fig. 2(b) presents an evolution of $X$-parameter with Pr concentration. In general, six-fold coordinated Ir ions in IrO$_{6}$ octahedra have equal Ir-O bond-length ($d_{Ir-O}$). The $X$-parameter for an undistorted IrO$_{6}$ octahedron turns out to be $X_{ideal}$ = 0.3125 which generates a perfect local cubic crystal field. The deviation of X from its ideal value generates a trigonal crystal field which modifies the orbital ordering or local hybridization.\cite{JP,Hozo} The $X$ value (0.36) for Y$_2$Ir$_2$O$_7$ implies IrO$_{6}$ octahedra are distorted which lowers the symmetry by compressing the octahedra. Fig. 2b depicts that $X$-parameter decreases over the series which is suggestive of decreasing trigonal distortion with Pr. This variation of $X$-parameter is though directly associated with the $<$Ir-O-Ir$>$ bond angle and the Ir-O bond-length, hence hopping of charge carriers. Figs. 2c and 2d show the change of Ir-O bond length and $<$Ir-O-Ir$>$ bond-angle, respectively with Pr doping. The Ir-O bond-length decreases while $<$Ir-O-Ir$>$ bond-angle increases with $x$. This implies that with Pr doping, the distortion in IrO$_{6}$ octahedra reduces and the Ir-5$d$/O-2$p$/Ir-5$d$ orbital overlapping increases which would facilitates the hopping of itinerant Ir-5$d$ electrons.

\subsection{X-ray absorption spectroscopy study}
The x-ray absorption spectroscopy (XAS) measurements have been done on selected samples ($x$ = 0.0, 0.4 and 1.0) of present series to understand the cationic charge state and the Ir-O hybridization state. The $L_{2}$ (2$P_{1/2}$ $\rightarrow$ 5$d$) and $L_{3}$ (2$P_{3/2}$ $\rightarrow$ 5$d$) absorption edge for Ir$^{4+}$ (5$d^5$) electronic state in XAS occur at 12.824 keV and 11.220 keV of energy, respectively.\cite{Bjkim,Clancy1} While the $L_{2}$ edge is mainly related to transitions of 5$d_{3/2}$ holes, the $L_{3}$ absorption edge involves transition of both 5$d_{3/2}$ and 5$d_{5/2}$ states.\cite{Clancy1,liu} Therefore, the relative intensity of $L_{2}$ and $L_{3}$ edge (so called statistical branching ratio) gives an information about ground state expectational value of the SOC $<$\textbf{L.S}$>$ of 5$d$ states. In present series, the SOC effect is unlikely to change, hence we have focused only on $L_{3}$ absorption edge spectra.

Fig. 3a shows normalized XAS spectra collected at $L_{3}$ absorption edge for present series (data for $x$ = 0.4 and 1.0 are shifted vertically). As evident in Fig. 3a, the $L_{3}$ edge occurs at 11.220 keV, which is in agreement with other Ir-based materials (i.e., Na$_2$IrO$_3$, Sr$_2$IrO$_4$ and Y$_2$Ir$_2$O$_7$).\cite{Bjkim,Clancy1,liu} We find that the position of $L_{3}$ edge shifts toward lower energy with progressive substitution of Pr (i.e., 11220.24, 11219.99 and 11219.91 eV for $x$ = 0.0, 0.4 and 1.0, respectively). The change in $L_{3}$ peak position is small ($\Delta$E $\sim$  -0.33 eV) over the series where the left inset of Fig. 3a shows a magnified view of absorption peak. We understand this change in $L_{3}$ peak position is due to change in Ir charge state ratio. Our recent x-ray photoemission spectroscopy (XPS) study shows that Ir$^{4+}$ is a major component while a small amount ($\sim$ 5.4\%) of coexisting Ir$^{5+}$ ions have been observed in Y$_2$Ir$_2$O$_7$.\cite{kumar} The small shifting of $L_{3}$ peak toward lower energy is possibly due to removal of Ir$^{5+}$ ions with Pr substitution. To further understand this, we have plotted the second derivative of $L_{3}$ spectra in right inset of Fig. 3a. As evident in figure, the derivative shows a shoulder on both sides of the main peak for $x$ = 0.0 while this shoulder disappears for the doped materials ($x$ = 0.4 and 1.0). The shoulder in second derivative is considered to be an indicative of coexisting Ir$^{5+}$ ions, which disappears with Pr substitution.\cite{laguna}

Further, to understand the unoccupied Ir-$d$ states or hybridization between the Ir-$d$ and O-2$p$ states, we have carried out XAS is measured at O-K edge at room temperature. Due to crystal field effect, Ir-$d$ orbitals are split into low lying $t_{2g}$ ($d_{xy}$, $d_{xz}$ and $d_{yz}$) and $e_g$ ($d_{x^2-y^2}$ and $d_{z^2}$) states where a strong SOC further split $t_{2g}$ state into $J_{eff}$ = 3/2 quartet and $J_{eff}$ = 1/2 doublet. In IrO$_{6}$ octadedra, orbitals ($p_x$, $p_y$ and $p_z$) of six ligand oxygen’s (four basal or in-plane and two apical) hybridize with these $d$ orbitals. The modification of local structural parameters (Fig. 2) would change the hybridization with Pr substitution. For $t_{2g}$, while $d_{xy}$ hybridizes only with basal $p_x$/$p_y$ the $d_{xz}$ and $d_{yz}$ participate with both basal $p_z$ and apical $p_x$/$p_y$ orbitals. In case of $e_g$ state, the $d_{x^2-y^2}$ and $d_{z^2}$ orbital hybridize with $p_x$/$p_y$ and $p_z$, respectively.\cite{liu,kang,brouet} Fig. 3b shows normalized O-K edge spectra for $x$ = 0.0, 0.4 and 1.0 materials of present series. As evident in figure, two distinct peaks (marked by vertical dotted lines) in the low energy regime are seen at binding energies ($E_b$) 528.95 and 532.77 eV while a broad hump is observed between energy 534 and 548 eV. The energy gap between the first two peaks at low energy side is around 3.8 eV. In low energy side, the observed peak at 528.95 eV originates due to hybridization between Ir $d_{xz}$/$d_{yz}$ and apical O $p_x$/$p_y$ orbitals while the peak at 532.77 eV arises due to bonding between Ir-$t_{2g}$ with basal O-2$p$ orbitals. The broad hump at higher energy between 534 and 548 eV corresponds the transition to hybridized states between Ir $e_g$ and O 2$p$ orbitals. As seen in Fig. 3b, the peak position at 528.95 eV does not show any significant change but that at 532.77 eV shifts toward lower energy with Pr amount. For $x$ = 0.4 and 1.0 material, we find that peak at 532.77 eV changes by $\sim$ 0.56 and 0.63 eV, respectively. The shifting of second peak toward low energy side implies an increased hybridization of Ir-$t_{2g}$ orbitals with basal O-2$p$ orbitals which mostly arises due to larger size of Pr$^{3+}$ions. Here, we note that the present analysis is based on model of isolated $t_{2g}$ and $e_g$ orbitals. However, a recent theoretical calculation has shown the effect of CFE, SOC and $U$ on mixing of $t_{2g}$ and $e_g$ orbitals where the effect becomes prominent in case of 5$d$ transition metals.\cite{stamo} While the low-energy peaks and high-energy hump in XAS spectra are attributed to the hybridization of Ir-$t_{2g}$/O-$p$ and Ir-$e_g$/O-$p$ orbitals (Fig. 3b and above discussion), respectively the positions of peaks/hump are seen to vary across the composition and crystal structure.\cite{liu,kang,brouet,sohn} Further, O-K edge spectra in Fig. 3b show prominent shoulder in peaks/hump. Therefore, theoretical calculations are required to understand the possible role of $t_{2g}$ and $e_g$ orbital mixing on XAS spectra of Ir based materials.

\begin{figure}
	\centering
		\includegraphics[width=6.2cm]{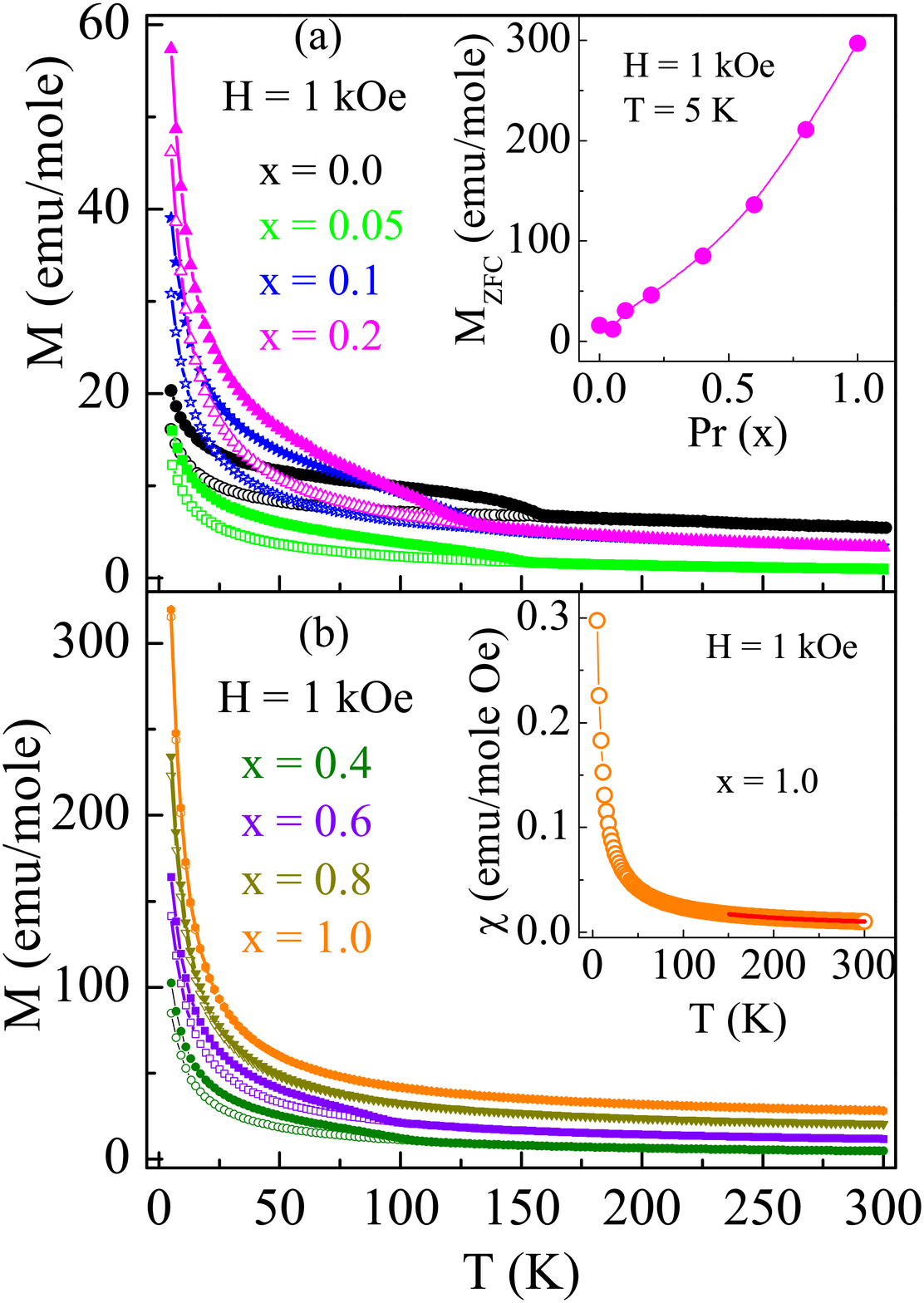}
	\caption{ Temperature dependent magnetization data measured in 1 kOe under ZFC and FC protocol have been shown for (Y$_{1-x}$Pr$_x$)$_2$Ir$_2$O$_7$ series with (a) $x$ = 0.0. 0.05, 0.1, 0.2 (b) $x$ = 0.4, 0.6, 0.8, 1.0 samples. The data in Fig. 4 (b) are vertically shifted with 5, 12 18 emu/mole for $x$ = 0.6, 0.8, 1.0. Inset of (a) shows $M_{ZFC}$ value at 5 K with Pr level. The inset of (b) shows the magnetic susceptibility $\chi$ = $M/H$ for Pr$_2$Ir$_2$O$_7$ ($x$ = 1.0) where the solid line in high temperature regime is due to Eq. 1.}
	\label{fig:Fig4}
\end{figure}

We have also attempted to understand the valence state of Pr which is important for its role in governing the electronic and magnetic properties. In Pr$_6$O$_{11}$ (which is used as ingredient for material synthesis), Pr exists in mixed valence state of 3+ (Pr$_2$O$_3$) and 4+ (PrO$_2$). The measured $M_{4}$ and $M_5$ edge absorption spectra for doped $x$ = 0.4 and 1.0 samples of present series are shown in Fig. 3c. As evident in figure, the line shape and peak position of $M_{4}$ and $M_5$-edge absorption spectra are similar for both these materials, however, the intensity of the $M_{4,5}$ increases with Pr concentration. The position of $M_5$ and $M_4$ peaks is found to be around 930 and 949.3 eV, respectively which closely match with that for Pr$^{3+}$ ions while corresponding peak positions for Pr$^{4+}$ occurs at higher energy.\cite{martin,hu} This suggests that Pr is in 3+ charge state, thus shows an agreement with Ir$^{4+}$ charge state.

\subsection{Magnetization study}
Fig. 4. shows the temperature ($T$) dependent magnetization data ($M$) for (Y$_{1-x}$Pr$_x$)$_2$Ir$_2$O$_7$ series where the data have been collected following zero field cooled (ZFC) and field cooled (FC) protocol in an applied magnetic field of 1 kOe. It is evident in Fig. 4a that $M(T)$ for Y$_2$Ir$_2$O$_7$ exhibits a magnetic irreversibility between ZFC and FC magnetization around $T_{N}$ = 160 K, opening a gap below this temperature.\cite{harish,kumar} This is considered to be an onset temperature for long range AFM ordering. The $M_{ZFC}$, however, do not show any prominent cusp/peak around $T_{N}$. Here, it can be mentioned that the low temperature magnetic state in Y$_2$Ir$_2$O$_7$ is quite intriguing as muon spin rotation ($\mu$SR) and neutron powder diffraction (NPD) experiments have shown an opposite result in case of long-range magnetic ordering.\cite{shapiro,disseler} Recently, we have demonstrated that the low temperature magnetic state in Y$_2$Ir$_2$O$_7$ as well as in its doped materials Y$_2$(Ir$_{1-x}$M$_2$)$_2$O$_7$ (M = Ru and Ti) has nonequilibrium ground state.\cite{harish,kumar,hkumar} The effect of Pr doping on magnetization is shown in Figs. 4a and 4b. The notable observation is that except for $x$ = 0.05, magnetic moment at low temperature continuously increases with $x$ (inset of Fig. 4a). This increase of magnetic moment indicates a weakening of magnetic exchange interaction where the spins are unlocked from AFM interaction, hence yielding a higher moment. The transition temperature $T_{N}$, on the other hand, shows a continuous decrease with $x$ where for Pr$_2$Ir$_2$O$_7$ ($x$ = 1.0), the $M_{ZFC}$ and $M_{FC}$ magnetization merge completely which implies a PM-like behavior.\cite{nakatsuji}

For high temperature PM state, the magnetic susceptibility ($\chi$ = (M/H) has been analyzed using modified
Curie-Weiss law,

\begin{eqnarray}
	\chi = \chi_0 + \frac{C}{T - \theta_P}
\end{eqnarray}

where $\chi_0$, $C$, and $\theta_P$ are the temperature independent magnetic susceptibility, the Curie constant, and Curie-Weiss temperature, respectively. A representative fitting of $\chi(T)$ data with Eq. 1 in temperature range between 150 - 300 K is shown in inset of Fig. 4b as solid line for $x$ = 1.0. A reasonably good fitting with Eq. 1 implies magnetic state in PM state obeys Curie-Weiss behavior. We obtain fitting parameters $\chi_0 = 2.3\times10^{-3}$ emu/mole, $C$ = 2.58(1) and $\theta_P$ = -22.8(4) K.\cite{nakatsuji} Similar good fitting has been obtained for remaining samples in present series. The corrected inverse susceptibility ($\chi$ - $\chi_0$)$^{-1}$ vs $T$ shows a linear behavior in high temperature regime (Fig. 5). Along with $T_{N}$, the composition dependent $\theta_P$ are shown in Fig. 6a. While $T_{N}$ shows a linear decrease till $x$ = 0.8 the $\left|\theta_P\right|$, on the other hand, shows a steep increase till $x$ = 0.2 and then changes are not significant. The obtained values of both $T_{N}$ and $\left|\theta_P\right|$ for $x$ = 0.0 and 1.0 materials agree with the reported values.\cite{nakatsuji,harish,kumar} Here, it can be noted that for Pr$_2$Ir$_2$O$_7$, even though $T_{N}$ vanishes but it exhibits a finite $\left|\theta_P\right|$. This finite $\left|\theta_P\right|$ arises due to Ir-5$d$ electrons mediated AFM-type RKKY interactions among Pr-4$f$ atoms.\cite{nakatsuji} Here, we put a word of caution that magnetic parameters ($\theta_P$, $f$ and $\mu_{eff}$) in Fig. 6 are determined by fitting Eq. 1 in temperature range from 150/200 K to 300 K. Though this temperature range is well with PM state but this is indeed a limited temperature range, and one needs to check this fitting in an extended high temperature range for a correct value. Nonetheless, parameters are found to be consistent with each other and agree with the previous reports.    

\begin{figure}
	\centering
		\includegraphics[width=8cm]{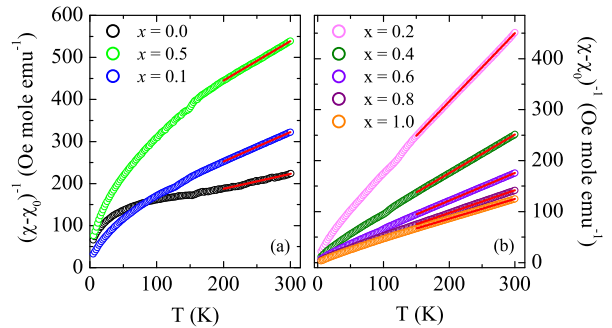}
	\caption{Temperature dependent inverse susceptibility ($\chi$ - $\chi_0$)$^{-1}$ are shown for (Y$_{1-x}$Pr$_x$)$_2$Ir$_2$O$_7$ series with (a) $x$ = 0.0. 0.05, 0.1 (b) $x$ = 0.2, 0.4, 0.6, 0.8, 1.0 composition. Solid lines are due to straight line fitting in high temperature regime.}
	\label{fig:Fig5}
\end{figure}

The long-range magnetic ordering is, however, suppressed in Pr$_2$Ir$_2$O$_7$, due to Kondo screening of Pr moments. We have further calculated the frustration parameter $f$ ($\left|\theta_P\right|$/$T_{N}$) for whole series, as shown in Fig. 6b. This parameter provides a vital information about a frustrated magnetic system where for $f$ = 1 the system realizes magnetic ordering temperature according to its exchange interaction strength. Here, it can be mentioned that frustration parameter mainly applies to isotropic systems, and $f$ much higher than 1 (usually, $>$ 10) for a system is considered to be a frustrated system. In Fig. 6b, the $f$ parameter remains small for present series, however, we have used this information to understand the evolution of magnetic nature in present series. The $f$ value for Y$_2$Ir$_2$O$_7$ is calculated as 2.07, showing a some level of frustration present in system. For low concentration of Pr, the $f$ decreases sharply and shows value $\sim$ 1 for $x$ = 0.1. The $f$ $<$ 1 for $x$ $>$ 0.1 is quite intriguing which indicates a low value of $\left|\theta_P\right|$ compared to its magnetic transition temperature $T_{N}$. This sudden suppression of $\left|\theta_P\right|$ occurs with the introduction of magnetic Pr$^{3+}$ in the place of nonmagnetic Y$^{3+}$ where a further strengthening of exchange interaction would have been a possible scenario with $x$. This is quite surprising because a reasonably strong AFM-type exchange interaction and transition temperature ($T_N$ $\sim$ 160 K) in Y$_2$Ir$_2$O$_7$, which is mainly due to Ir sublattice, is largely suppressed with an inclusion of magnetic Pr$^{3+}$ while Ir sublattice remains unaltered. We believe that this weakening of $\left|\theta_P\right|$ is due to $f$-$d$ interaction between localized Pr-4$f$ and itinerant Ir-5$d$ electrons. A RKKY type $f$-$d$ exchange interaction driven finite $\theta_P$ has been shown for Pr$_2$Ir$_2$O$_7$.\cite{nakatsuji} The effect of $f$-$d$ exchange interaction for pyrochlore iridates, in general, has been discussed in a recent study.\cite{chen} The $f$ for Pr$_2$Ir$_2$O$_7$ could not be calculated due to unavailability of $T_{N}$ but it has been reported to be $\sim$ 170 implying it is a highly frustrated spin-liquid material.\cite{nakatsuji} Nonetheless, our results conclusively show a disappearance of magnetic state with doping in (Y$_{1-x}$Pr$_x$)$_2$Ir$_2$O$_7$ series where the observed behavior has the similarity of quantum phase transition (QPT). The prominent example of doping induced QPT is Sr$_{1-x}$Ca$_x$RuO$_3$ where the system shows vanishing of magnetic state around 70\% of Ca doping.\cite{fuchs} In present series, magnetic state disappears at more than 80\% of Pr substitution. This calls for a detailed theoretical and experimental investigation as the quantum criticality has already been hinted by calculation for this series.\cite{lucile}    

\begin{figure}
	\centering
		\includegraphics[width=6cm]{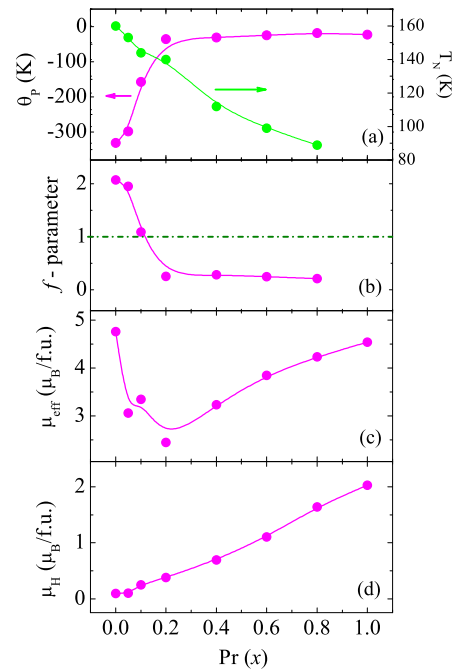}
	\caption{(a)  Curie-Weiss temperature $\theta_P$ and $T_N$ (b) Frustration parameter $f$ (c) Effective paramagnetic moment ($\mu_{eff}$) and (d) Moment $\mu_H$ at field 70 kOe are shown with doping concentration $x$ in (Y$_{1-x}$Pr$_x$)$_2$Ir$_2$O$_7$ series. These parameters are obtained from fitting of magnetization data with Eq. 1.}
	\label{fig:Fig6}
\end{figure}

An effective PM moment ($\mu_{eff}$) has been calculated from obtained Curie constant $C$. As evident in Fig. 6c, $\mu_{eff}$ initially decreases and then increases showing a dip around $x$ = 0.2. The spin-only $\mu_{eff} (= g \sqrt{S(S + 1)} \mu_B$ where g = 2 is the Lande g factor and S = 1/2) for Y$_2$Ir$_2$O$_7$ has been calculated with value 3.46 $\mu_B$/f.u. (2 Ir$^{4+}$ ions per f.u.) which is lower than the experimentally obtained value 4.76 $\mu_B$/f.u. (Fig. 6c). For the doped materials, the $\mu_{eff}$ can be calculated using 2$\sqrt{\left[\mu^{Ir}_{eff}\right]^2 + \left[(1 - x)\mu^{Y}_{eff}\right]^2 +  x\left[\mu^{Pr}_{eff}\right]^2}$. Given that $\mu^{Y}_{eff}$ does not contribute and $\mu^{Pr}_{eff}$ = 5.65 $\mu_B$/f.u., the expected $\mu_{eff}$ for whole system will increase with $x$. However, the initial decrease of $\mu_{eff}$ in present series is probably due to the presence of local AFM-type Pr-Ir interaction in PM state which diminishes for $x$ $>$ 0.2. A lowering of moment has also been observed in $M(T)$ data in low doped materials in Fig. 4a.

\begin{figure}
	\centering
		\includegraphics[width=8cm]{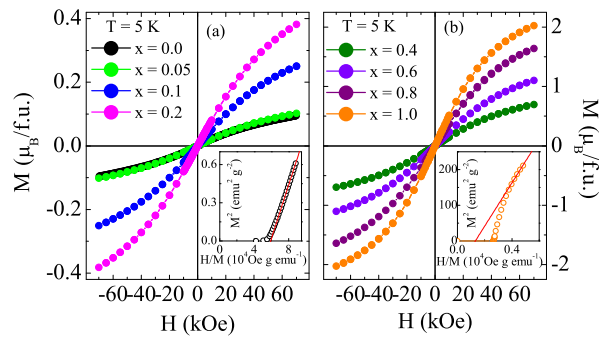}
	\caption{Magnetic field dependent magnetization are shown for (Y$_{1-x}$Pr$_x$)$_2$Ir$_2$O$_7$ series with (a) $x$ = 0.0. 0.05, 0.1, 0.2 (b) $x$ = 0.4, 0.6, 0.8, 1.0 composition at 5 K. Inset of (a) and (b) shows the Arrott plot of the $M(H)$ data for $x$ = 0.0 and 1.0, respectively where the solid lines are due to straight line fitting.}
	\label{fig:Fig7}
\end{figure}

Magnetic field dependent magnetization data $M(H)$ collected at 5 K are shown in Figs. 7a and 7b for (Y$_{1-x}$Pr$_x$)$_2$Ir$_2$O$_7$ series. For Y$_2$Ir$_2$O$_7$, $M(H)$ data is non-linear with no sign of saturation till highest measuring field of 70 kOe. With progressive substitution of Pr, major observations are the nonlinearity in $M(H)$ increases and the magnetic moment $\mu_H$ at 70 kOe shows a continuous increases (Fig. 6d). A close observation reveals $M(H)$ for Y$_2$Ir$_2$O$_7$ show a small magnetic coercivity $H_c$ ($\sim$ 100 Oe) and a remnant magnetization $M_r$ ($\sim$ 4.7$\times10^{4}$ $\mu_B$/f.u). The $H_c$ almost vanishes with $x$. The $M(H)$ data are further plotted in form of Arrott’s plot\cite{arrott} ($M^2$ vs $H/M$) for two end members i.e., $x$ = 0.0 and 1.0, respectively. A positive intercept due to straight line fitting in Arrott’s suggests a spontaneous magnetization or ferromagnetic (FM) behavior.\cite{kumar} As seen in insets of Fig. 7, both the intercepts are negative which is indicative of a non-FM state. As evident in figure, the magnitude of intercept decreases with Pr which is in good agreement with decrease of $\theta_P$ (Fig. 6a).

It can be here mentioned that similar hysteresis in $M(H)$ data of Y$_2$Ir$_2$O$_7$ has previously been observed by Zhu \textit{et al.}\cite{zhu} which authors attribute due to coexisting (weak) FM phase on top of (large) AFM background. However, it would be difficult to conclude about coexisting FM phase from such small hysteresis loop alone. Further, such a weak FM phase will unlikely be reflected in Arrott plot.

\begin{figure}
	\centering
		\includegraphics[width=8cm]{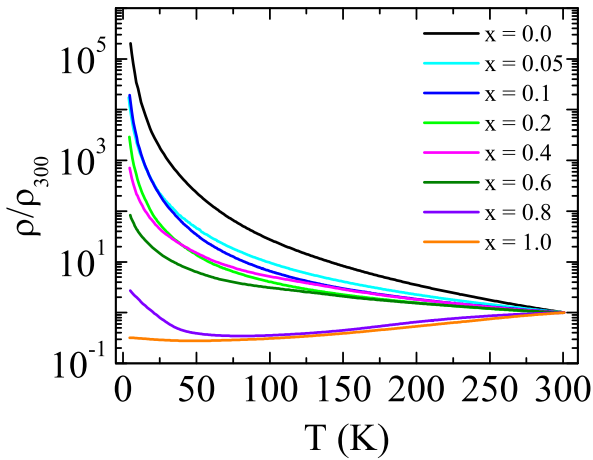}
	\caption{Temperature dependent resistivity normalized with $\rho_{300}$ are shown for (Y$_{1-x}$Pr$_x$)$_2$Ir$_2$O$_7$ series.}
	\label{fig:Fig8}
\end{figure}

\subsection{Electrical transport}
The temperature dependent resistivity $\rho$(T) for (Y$_{1-x}$Pr$_x$)$_2$Ir$_2$O$_7$ series are shown in Fig. 8 after normalizing with its value at 300 K i.e., $\rho_{300}$. The Y$_2$Ir$_2$O$_7$ shows an insulating behavior where the resistivity increases by couple of orders in going to low temperatures.\cite{disseler,harish,kumar} It is evident in figure that $\rho$/$\rho_{300}$ decreases continuously with substitution of Pr while Pr$_2$Ir$_2$O$_7$ shows a metallic charge conduction.\cite{nakatsuji} In fact, resistivity at low temperature changes drastically by about 10 orders over the series. Along with a PM state, the arising of a metallic state in robust AFM-insulating state of Y$_2$Ir$_2$O$_7$ with substitution of magnetic Pr$^{3+}$ is quite intriguing. However, the other pyrochlore iridates A$_2$Ir$_2$O$_7$ with magnetic A-ions (i.e., Nd$^{3+}$, Sm$^{3+}$, Gd$^{3+}$) elements show a metal-insulator transition (MIT) with an insulating state at low temperature.\cite{Matsuhira} In addition to electron doping with its 4$f^2$ structure, Pr$^{3+}$ has comparatively higher ionic radii which would induce a local structural modifications that will influence the Ir-O hybridization. Indeed, an increased hybridization between Ir-$t_{2g}$ and (basal) O-2$p$ orbitals has already been observed in Pr doped materials (Fig. 3b). This will facilitate the charge conduction, hence a metallic state is expected. For $x$ = 0.8, the $\rho (T)$ shows a MIT around 81 K which closely matches with its magnetic transition temperature $T_N$ (Fig. 6a). Although Pr$_2$Ir$_2$O$_7$ ($x$ = 1.0) shows a metallic behavior but an upturn in its $\rho(T)$ is seen around 50 K which is in agreement with previous study.\cite{nakatsuji} The nature of electron conduction in insulating samples ($x$ $\leq$ 0.6) has been analyzed with following power-law behavior,

\begin{eqnarray}
	\rho = \rho_0 T^{-n}
\end{eqnarray}

where $n$ is an exponent. Fig. 9a shows $\rho(T)$ data in $\log-\log$ scale where the straight lines are due to fitting of data with Eq. 2. For Y$_2$Ir$_2$O$_7$, the exponent $n$ has been obtained to be 2.98 which decreases with $x$ (inset of Fig. 9a). The semi-$\log$ plotting of $\rho(T)$ data for $x$ = 0.8 and 1.0 materials show a dip where a $\ln T$ dependence at low temperature implies a Kondo-like behavior (Figs. 9b and 9c).

A Kondo-like behavior has been indicated for Pr$_2$Ir$_2$O$_7$ ($x$ = 1) at low temperature.\cite{nakatsuji} For the compound with $x$ = 0.8, there are two important observations, $\rho(T)$ follows a $\ln T$ dependence at low temperature (Fig. 9b) and the dip in $\rho(T)$ closely matches with its $T_N$ (Fig. 8). Therefore, this dip in $\rho(T)$ may be either due to Kondo-like behavior as in $x$ = 1 or metal-insulator transition (MIT). This dip temperature ($\sim$ 81 K) appears too high for Kondo behavior, in particular when Pr-site is diluted with nonmagnetic Y$^{3+}$. Here, Kondo behavior arises due to $f$-$d$ interaction between localized Pr-4$f$ magnetic ions and itinerant Ir-5$d$ electrons. Therefore, it would be unlikely that the Kondo temperature increases in going from Pr$_2$Ir$_2$O$_7$ to 20\% diluted Pr lattice in $x$ = 0.8. On the other hand, many pyrochlore iridates (i.e., Nd$_2$Ir$_2$O$_7$, Sm$_2$Ir$_2$O$_7$, and Eu$_2$Ir$_2$O$_7$) have shown metal-insulator transition across its magnetic ordering temperature.\cite{Matsuhira} Given that dip temperature for $x$ = 0.8 is reasonably high ($\sim$ 81 K) and it closely matches with its $T_N$, the low temperature rise of $\rho(T)$ in this material is likely to be associated with the metal-insulator transition. 

\begin{figure}
	\centering
		\includegraphics[width=8cm]{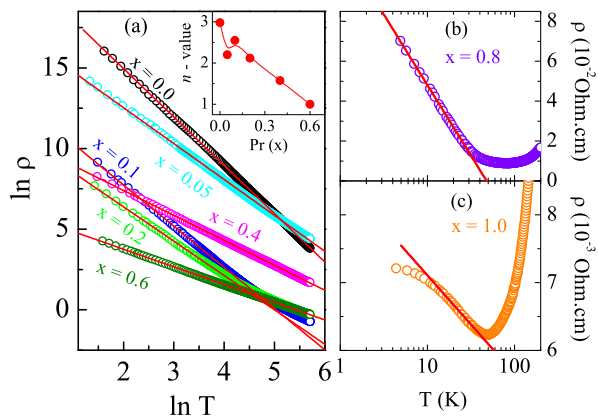}
	\caption{(a) The $\rho(T)$ data are plotted in ln-ln scale for (Y$_{1-x}$Pr$_x$)$_2$Ir$_2$O$_7$ series with $x$ = 0.0. 0.05, 0.1, 0.2, 0.4, 0.6. Inset shows the composition dependent exponent $n$ (Eq. 2) where the solid line is guide to eye. (b) and (c) show $\rho$ vs $\ln T$ form for $x$ = 0.8 and 1.0, respectively where the solid lines are due to straight fittings.}
	\label{fig:Fig9}
\end{figure}

The calculated magnetoresistance (MR), $\Delta \rho/\rho(0)$ = $\left[\rho(H) - \rho(0)\right]/\rho(0)$, for (Y$_{1-x}$Pr$_x$)$_2$Ir$_2$O$_7$ series at 2.5 K are shown in Fig. 10a. As evident in figure, the MR is not very large but their values are negative i.e., conductivity increases in presence of magnetic field. Moreover, MR does not show a systematic variation with $x$ but $x$ = 0.8 material shows a highest MR. Usually, negative MR in such materials is largely due to weak localization phenomenon, induced by quantum interference effect. Recently, a quadratic field dependence of negative MR has been seen at low temperature for Y$_2$Ir$_2$O$_7$.\cite{harish} Following that, MR data have been plotted as a function of $H^2$ in Fig. 10b. As seen in Fig. 10b, MR shows a $H^2$ dependence roughly above 25 kOe.

The insulating states in Ir-based pyrochlore materials are of general interest which have been investigated using different chemical substitutions at Ir-site but very limited studies are reported regarding A-site doping. In case of Pr$^{3+}$ substitution in (Nd$_{1-x}$Pr$_x$)$_2$Ir$_2$O$_7$, the resistivity decreases and the MIT in original system is suppressed for around 90\% doping of Pr.\cite{mat} In present case, the suppression of the insulating state with around 80\% of Pr imply a similar result. A transition from an insulating to metallic state in pyrochlore iridates appears to be related with ionic radii of $A$-site cation.\cite{Matsuhira} Here, Pr$^{3+}$ has larger ionic radii (1.126 \AA) compared to Y$^{3+}$ (1.019 \AA), therefore it would likely to enhance the electronic conductivity by reducing the trigonal distortion of IrO$_6$ octrahedra or equivalently increasing the Ir-O orbital hybridization.\cite{william,yang} For present $x$ = 0.8 sample (MIT temperature $\sim$ 81 K), we have calculated the average ionic radii to be 1.1054 \AA which almost matches with that of Nd$^{3+}$ (1.109 \AA). Here, note that Nd$_2$Ir$_2$O$_7$ exhibits a MIT around 33 K,\cite{Matsuhira} and this small difference in MIT temperature with our $x$ = 0.8 sample is due to slight difference in ionic radii.

\begin{figure}
	\centering
		\includegraphics[width=8cm]{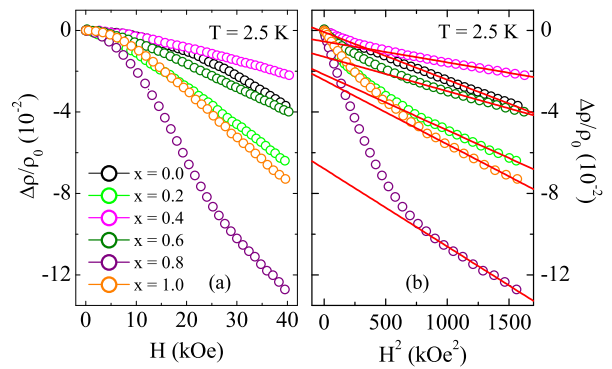}
	\caption{(a) shows MR as a function of magnetic field at 2.5 K for (Y$_{1-x}$Pr$_x$)$_2$Ir$_2$O$_7$ series. (b) shows the same MR as quadratic dependence of magnetic field. The solid lines are due to straight line fittings.}
	\label{fig:Fig10}
\end{figure}

The decrease of resistivity as well as MIT in present (Y$_{1-x}$Pr$_x$)$_2$Ir$_2$O$_7$ series is quite likely due to an increase of average A-site ionic radii. It can be noted that similar suppression of insulating state has previously been observed in Y$_2$Ir$_2$O$_7$ with Ca and Bi substitution at Y-site.\cite{fukazawa,soda,zhu,aito} The Ca$^{2+}$ (1.12 \AA) acts for hole doping in Y$_2$Ir$_2$O$_7$, and the complete suppression of insulating behavior occurs around 15\% of Ca doping.\cite{fukazawa,zhu} Similarly, Bi$_2$Ir$_2$O$_7$ shows complete metallic behavior where Bi$^{3+}$ (1.17 \AA) has slight higher ionic radii compared to Pr$^{3+}$.\cite{qi} A complete suppression of both insulating and magnetic state occurs around 50\% of Bi doping in Y$_2$Ir$_2$O$_7$ at Y-site.\cite{aito} The calculated average ionic radii of 50\% Bi doped sample is 1.097 \AA which is close to the value of our present 80\% Pr doped sample. Furthermore, it is previously observed that both MIT and magnetic phase for Eu$_2$Ir$_2$O$_7$ are suppressed with 10\% of Bi doping at Eu-site, however, the average ionic size is found to be 1.076 \AA.\cite{prachi} Moreover, it is observed that MIT across the $T_N$ in pyrochlore iridates occurs for materials which have average ionic size of $A$-site cations between 1.066 and 1.109 \AA.\cite{Matsuhira} Therefore, the observed minimum in $\rho(T)$ for $x$ = 0.8 is likely a MIT driven by magnetic ordering temperature. The prominent effect of $A$-site ionic radii on local structural parameters (trigonal distortion and Ir-O bond length/angle) in Fig. 2 indicate an increasing overlapping between of Ir-$d$ and O-2$p$ orbitals with Pr doping which has, indeed, been observed in spectroscopy measurements in Fig. 3. While the metal-insulating state in pyrochlore iridates is being seriously investigated, our results will hopefully shed light in this direction. However, in addition to orbital hybridization, any possible role of $f$-$d$ interaction on electronic transport needs to be understood using theoretical calculations.

\section{Conclusion}
In summary, a detailed structural, magnetic, and transport properties are studied in series of polycrystalline samples (Y$_{1-x}$Pr$_x$)$_2$Ir$_2$O$_7$ with $x$ $\leq$ 1.0. Structural analysis shows the system retains its original \textit{Fd$\bar{3}$m} cubic symmetry while the structural parameters show systematic changes with Pr. In particular, the distortion of IrO$_6$ octahedra decreases with $x$. The doped Pr increases overlapping/hybridization between Ir-$t_{2g}$ and basal O-$p$ orbitals, as evident from XAS data. Magnetization measurements reveal that magnetic state is weakened with Pr substitution where Pr$_2$Ir$_2$O$_7$ shows complete paramagnetic behavior. The parent compound Y$_2$Ir$_2$O$_7$ is a strong insulator but the Pr substitution decreases the electrical resistivity in whole series. For $x$ = 0.8, a metal-insulator transition is observed around 81 K while the end member of this series Pr$_2$Ir$_2$O$_7$ shows metallic behavior in whole temperature range. The nature of charge conduction is found to follow the power-law behavior for $x$ $\leq$ 0.6 while a Kondo-like behavior is observed at low temperature for (Y$_{0.2}$Pr$_{0.8}$)$_2$Ir$_2$O$_7$ and Pr$_2$Ir$_2$O$_7$ materials. A whole series of samples exhibit a weak negative MR at low temperature which shows quadratic field dependence at higher field.

\section{Acknowledgment} 
We acknowledge UGC-DAE CSR, Indore for the magnetic measurements. We are thankful to DST-FIST and DSR-PURSE for the low temperature high magnetic field facility and for the financial supports. HK acknowledges UGC, India for BSR fellowship. Authors thank Profs. J-F. Lee,  J-M. Chen, and Shu-Chih Haw for their support in the XAS measurements.

\end{document}